# Pattern Propagation Speed in Synfire Chains with Excitatory-Inhibitory Couplings

Baktash Babadi

*School of Intelligent Systems, Institutes for Studies in Theoretical Physics & Mathematics*

baktash@ipm.ir

**Abstract**

The speed of firing pattern propagation in a synfire chain, composed of non-leaky integrate-and-fire neurons, and assuming homogenous connection delays, is studied. An explicit relation, relating the propagation speed to the connecting weights distribution and other network parameters, is derived. The analytic results are then checked with a computer simulation. When the network is fed with a fully synchronized input pattern, the pattern propagation speed is independent of the weight parameters. When the fed input is asynchronous, depending on the weight parameters, the propagation speed is more than or less than the synchronous case. In this case the propagation speed increases by increasing the mean or standard deviation of connecting weights. The biological relevance of these findings and their relevance to the notion of synfire chains are discussed.

## 1. Introduction

Recent evidences from multiunit recording studies indicate that the cortical neural populations are able to produce precise firing sequences (Abeles 1993a,b ; Prut et al 1998). In a precise firing sequence, when the first neuron fires, the second neuron fires $t_1$ milliseconds later and the third neuron $t_2$ milliseconds later, with a sub-millisecond precision across trials.

The synfire chain model, proposed by Abeles (1991) on the basis of anatomical and physiological characteristics of the cerebral cortex, appeared to be able to explain the generation of precise firing sequences. A synfire chain consists of successive pools of spiking neurons connected in a feed-forward manner. If all neurons in one pool fire synchronously, they will in turn elicit a synchronous firing pattern in the next pool and a train of synchronous firing patterns will be generated in the successive pools. In this case, when one observes three different neurons from three successive pools, he will detect a precise firing sequence among them. Given that all neurons in a pool are synchronous, the same precise firing sequence can be detected in several observations.

To date, researchers have analyzed different aspects of the synfire chain models. These aspects include memory capacity (Herrmann et al 1995), learning algorithms (Hertz & Prügel-Bennett 1996a,b; Sougné 2001) and robustness of synchronous pattern propagation in spite of noise (Postma et al 1999; Diesman et al 1999; Gewaltig et al 2001) etc.

Assuming that a synfire chain is responsible for generating the observed precise firing sequences, the question arise that what properties of the synfire model are most important in this regard?

As described earlier in the literature (Diesman et al 1999; Gewaltig et al 2001; Cateau & Fukai 2001), an important aspect of synfire activity is the sustainability of synchronous firing patterns through the successive pools.

As far as its role in generation of precise firing sequences is concerned, another important aspect of synfire activity is the speed of pattern propagation from one pool to the next. Due to the synfire notion, it is the velocity of pattern propagation through the successive neuron pools that determines the time interval between single spikes within a recorded firing sequence.

Pattern propagation velocity has been addressed in some studies using numerical simulations (Arnoldi & Brauer 1996) and analytical methods (Arndt et al 1995; Wennekers & Plam 1996; Wennekers 2000). Arnoldi and Brauer (1996) have shown that two parallel synfire chains with different propagation speeds, coupled with excitatory synapses, will become synchronized eventually. Wennekers and Plam (1996) and also Wennekers (2000) have shown that the speed of pattern propagation in synfire chains considerably depends on the non-specific external input (background activity) and also the number of co-activated chains. In these studies, the analytical approach was restricted to over-simplified models, which do not allow the network properties such as connecting weights distribution and temporal spread of patterns in each pool to be taken into account. On the other hand, the more detailed and biology inspired models were tackled only by means of computer simulations (Arnoldi & Brauer 1996), which can hardly capture the behavior of the model holistically.

In this article we have extended the analytical method used by Herrmann et al (1995), to account for propagation speed in synfire chains. Thereby, we have derived an explicit formula, relating the speed of pattern propagation between two successive layers, to the network parameters. In order to incorporate the effect of firing pattern temporal jitter into the analysis, we have also used the notion of pulse packets, which was introduced by Diesmann et al (1999).

Given that the most learning algorithms proposed for the synfire chains (Hertz, Prügel-Bennett 1996a,b; Sougné 2001), are based on the modification of connection weights, we focused our concentration on the effect of connection weights distribution parameters on the propagation speed. However, it is worth to note that our analytical method can be likewise used to study the effect of other network parameters.

In the next section the structure of our network model is introduced. In section 3, we have applied the analytical methods to study the speed of pattern propagation in the network. In section 4, the analytical results are verified with a computer simulation of the network. Finally in section 5, the biological relevance of the results, particularly to the synfire model as a structure involved in the generation of precise firing sequences is discussed.

**2. The network model**

The network studied in this article, is a multi-layered, feed-forward network, which consists of non-leaky integrate-and-fire neurons. All the layers contain equal number of neurons (namely *N*) (Figure 1.a).

**2.1 The neuron Model**

The non-leaky integrate-and-fire neuron model is a simplified approximation to the more detailed conductance based neuron models (Koch 1999). The voltage of a non-leaky integrate-and-fire neuron changes continuously as a function of its input with a rate adjusted to the real neuron's membrane time constant. Whenever this continuous changing potential reaches a certain value (the threshold), an action potential is artificially inserted, which simulates the firing event. Immediately after the action potential, the voltage will be reset to the resting potential. Although the non-leaky integrate-and-fire model is a caricature of a real neuron, it captures many of the real neuron properties and is much more realistic than the classic binary (Ising) neuron models in which the states of the neuron are discrete.

The resting membrane potential of this neuron is assumed to be zero and in the sub-threshold regime, the membrane potential is governed by:

$$\tau \frac{dV_i^m}{dt} = Input_i^m(t),  \qquad (1)$$

Where, $V_i^m$ is the voltage of the *i*th neuron at layer $m$, $\tau$ is the membrane time constant of the neuron, which is taken equal for all the neurons in the network, and $Input_i^m(t)$ is the incoming input to the *i*th neuron at layer $m$.

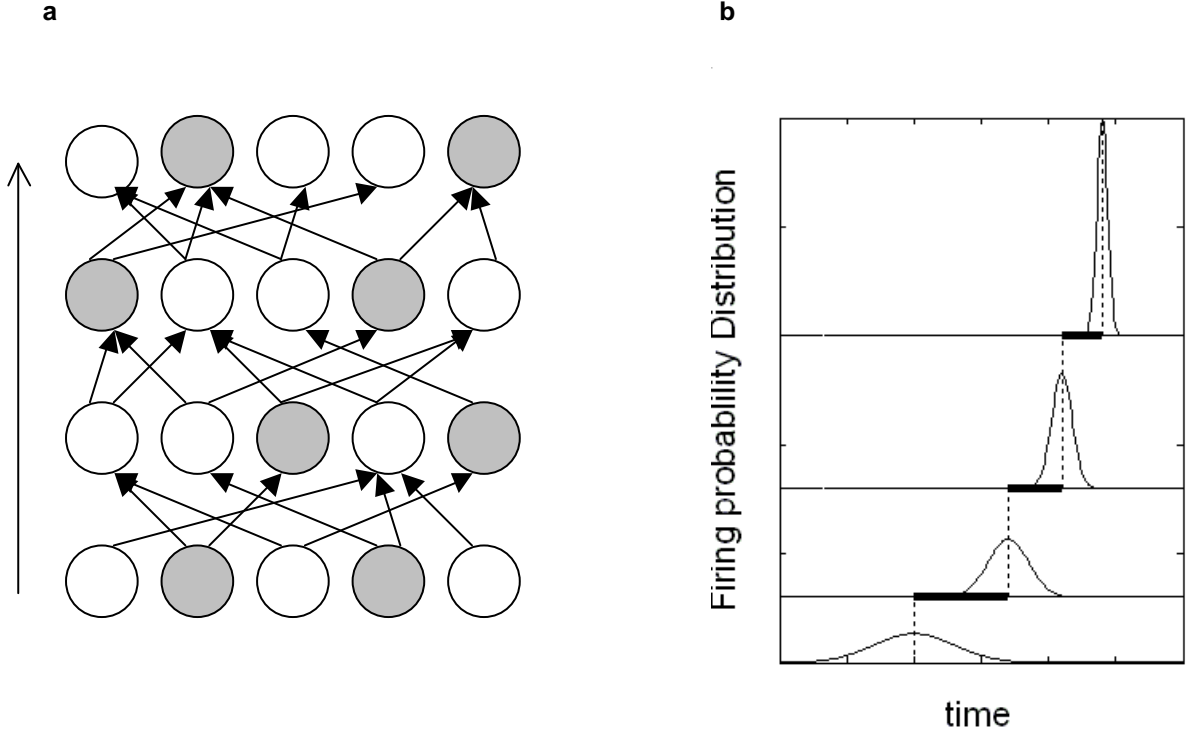

**Figure 1. a)** A schematic view of the studied network. **b)** Schematic view of the probability density of firing times for neurons is successive layers. The mean-to-mean (or peak-to-peak) interval (the bold lines) is an index for pattern propagation delay between successive layers.

### 2.2 The Input

The neurons in the first layer, which are the initiators of synfire activity, are assumed to fire at independent, normally distributed random times. The temporal probability density function for each pulse generated in the first layer can be written as:

$$r_1(t) = \frac{1}{\sqrt{2\pi} s_1} Exp(-0.5(\frac{t}{s_1})^2) \qquad (2),$$

where, $s_1$ is the temporal standard deviation of the pulses. Without loss of generality, we assumed that the mean of firing times for the neurons in the first layer is zero. Thus, according to the pulse packet notion (Diesmann et al 1999), the first layer produces a pulse packet containing $n_1 = N$ pulses with a temporal standard deviation of $s_1$.

Except for the first layer, the input of each neuron is the weighted sum of the outputs of neurons in the previous layer, as illustrated in the following formula:

$$Input_i^m(t) = \sum_j w_{ij}^m \, Output_j^{m-1}(t - t_{ij}^d),  \qquad (3)$$

in which $w_{ij}^m$ is the weight of the connection between the *j*th neuron of the donor layer (layer *m*-1) to the *i*th neuron of the recipient layer (layer *m*), $Output_j^{m-1}$ is the spike generated by the *j*th neuron of donor layer (layer *m*-1), and $t_{ij}^d$ is the connection delay. The delays for all connections are assumed to be equal, so we can omit the indices *i,j* and simply denote the delay by $t^d$ instead.

### 2.3 The connections

Two successive layers are fully interconnected but there is no connection within a layer itself. There is also no feed back connection from one layer to the previous layers. All synaptic weights of the network are constant and distributed normally with mean weight of $\overline{w}$ and standard deviation of $s_w$.

### 2.4 Output

A constant threshold value (*th*) is assigned to each neuron, which does not vary in time. The thresholds of all neurons are assumed to be equal. Whenever the membrane voltage of a neuron reaches its threshold, it fires, i.e. generates a spike which can be represented mathematically by a delta (impulse) function as:

$$Output_j^m(t) = \delta(t - t_j^m), \qquad (4)$$

where, $t_j^m$, is the firing time of the neuron, i.e. the time when its voltage has reached the threshold. Immediately after firing, the voltage of the neuron is rendered zero. In real neurons the duration of an effective action potential is considerably short, also immediately after producing an action potential the membrane potential declines, so the above assumptions on the firing state

(namely setting voltage to zero and using $\delta$ (impulse function) as the ultimate output function) seem to be relevant. In this analysis, we assume that each neuron produces one spike at the most in response of the incoming pulse packet.

As mentioned above, the firing times of the neurons in the first layer have a distribution with mean $\overline{t_1} = 0$ and standard deviation $\sigma_1$. The neurons in the next layers in turn have firing times with some probability distribution characterized by a mean $(\overline{t_m})$ and a standard deviation $(\sigma_m)$, i.e. each layer $m$ produce a pulse packet containing $n_m$ pulses with a temporal standard deviation $\sigma_m$, centered around $\overline{t_m}$.

As most of the neurons in layer $m$ fire in times near $\overline{t_m}$, the time interval between $\overline{t_{m+1}}$ and $\overline{t_m}$ can be used as a measure for pattern propagation delay between layers $m+1$ and $m$. Thus, we use the mean-to-mean (or peak-to-peak) interval as an index for pattern propagation delay between successive layers (Figure 1.b). Note that the pattern propagation delay has an inverse relation with the propagation speed, i.e. the faster the propagation of pattern between to layers is, the shorter the propagation delay will be. In the rest of this article, we will focus our study on the effect of the network parameters on the pattern propagation delay.

## 2. Analytic method

Inserting eq.(3) and eq.(4) in eq.(1) and solving the differential equation for the $i$th neuron in layer $m+1$ yields:

$$V_i^{m+1} = \frac{1}{\tau} \sum_j w_{ij} \, \theta \, (t - t_j^m - t^d) \qquad (5)$$

Now assume that the firing times of the neurons in layer $m$ (the $t_j^m$ variables in (5)) obey a normal distribution:

$$r_m(t) = \frac{1}{\sqrt{2\pi}s_m} Exp(-0.5(\frac{t-\overline{t_m}}{s_m})^2) \tag{6}$$

As it can be seen in section 4, this assumption is a reasonable approximation in a wide range of network parameters.

With this assumption, when the number of firing neurons in the layer $m$ is sufficiently large, eq.(5) can be approximated as:

$$V_i^{m+1}(t) = \frac{n_m W_i}{t} \int_{-\infty}^{t} r_\mu(t' - t^d) dt' = \frac{n_m W_i}{t} \Phi(\frac{t - \overline{t_m} - t^d}{s_m}) \tag{7}$$

where $n^m$ is the number of firing neurons in layer $m$, $W_i = \sum_j \frac{w_{ij}}{n_m}$ is a measure for the average of incoming weights to the neuron and $\Phi(x)$ is the cumulative standard normal distribution. Given that the weights of the network obey a normal distribution with mean $\overline{w}$ and standard deviation $s_w$, the $W_i$ will obey a normal distribution with mean $\overline{w}$ and standard deviation $\frac{s_w}{\sqrt{n_m}}$ respectively. So, by eq.(7), it is obvious that $V_i^{m+1}(t)$ has also a normal distribution with the following parameters:

$$\begin{cases} \overline{V^{m+1}} = \frac{n_m \overline{w}}{t} \Phi(\frac{t - \overline{t_m} - t^d}{s_m}) \\ s_{V^{m+1}} = \frac{\sqrt{n_m} s_w \Phi(\frac{t - \overline{t_m} - t^d}{s_m})}{t} \end{cases} \tag{8}$$

as its mean and standard deviation.

Clearly, when $V_i^{m+1}(t)$ becomes greater than or equal with the threshold, the neuron fires. So the probability of firing before the time $t$ for the neurons in layer $m+1$ can be written as:

$$?_{m+1} = \int_{-\infty}^{u(t)} \frac{Exp(-\frac{x^2}{2})}{\sqrt{2p}} dx,$$

$$u(t) = \frac{th - \overline{V^{m+1}}}{s_{V^{m+1}}} \tag{9}$$

Evidently, the probability distribution of firing for the neurons will be:

$$r_{m+1}(t) = \frac{d?_{m+1}(t)}{dt} = \frac{du(t)}{dt} \frac{Exp(-\frac{u^2}{2})}{\sqrt{2p}} \tag{10}$$

Now, we must calculate the mean of $r_{m+1}(t)$ namely $\overline{t_{m+1}}$. Assuming that $\overline{t_{m+1}}$ is near $\overline{t_m} + t^d$, $u(t)$ can be approximated as:

$$u(t) \approx \frac{n_m \overline{w}(\sqrt{2p}s_m + 2t - 2\overline{t_m} - 2t^d) - 2\sqrt{2p}s_m th.t}{\sqrt{n_m}s_w(\sqrt{2p}s_m + 2t - 2\overline{t_m} - 2t^d)} \tag{11}$$

With this approximation, $\overline{t_{m+1}}$ can be calculated as:

$$\overline{t_{m+1}} = \overline{t_m} + t^d + \sqrt{2p}s_m(\frac{th.t(\sqrt{n_m^2 \overline{w}^2 + 8ns_w^2} - n_m \overline{w})}{4n_m s_w^2} - \frac{1}{2}) \tag{12}$$

and finally, the mean-to-mean interval for the layers $m+1$ and $m$, which is an index of pattern propagation delay, is:

$$D = \overline{t_{m+1}} - \overline{t_m} = t^d + \sqrt{2p}s_m(\frac{th.t(\sqrt{n_m^2 \overline{w}^2 + 8ns_w^2} - n_m \overline{w})}{4n_m s_w^2} - \frac{1}{2}) \tag{13}$$

which relates the mean-to-mean interval to the network parameters explicitly.

Note that when the firing pattern in the layer $m$ (the donor layer) is fully synchronous, $s_m$ will be zero, hence the second term in the right side of eq.(13) will be also zero. In this case the pattern propagation speed is independent of the network parameters except the connection delay $t^d$. When firing pattern of the layer $m$ is not fully synchronized ($s_m > 0$), the propagation speed can be lower than or higher than the fully synchronous case, i.e. if the second term in the right side of eq.(13) is positive, the mean-to-mean interval will be greater than $t^d$, hence the pattern propagation speed will be lower and vice versa. The condition for the second term in the right side of (13) to be positive is:

$$\overline{w} < \frac{2\,th\,\boldsymbol{t}}{n_m} - \frac{\boldsymbol{s}_w^2}{th\,\boldsymbol{t}} \tag{14}$$

From eq.(13) it is obvious that $\frac{dD}{dw} > 0$ and $\frac{dD}{d\boldsymbol{s}_w} > 0$.(provided that all the parameters are positive values). Thus, when the other parameters of the network are fixed, increasing the mean of connection weights will increase the propagation speed respectively, and increasing it more than a critical value, can cause the propagation speed become even more than the fully synchronized case.

Given that $\frac{dD}{d\boldsymbol{s}_w}$ is also always positive, a similar reasoning is also applicable for the weights standard deviation. In this regard, considering all the other parameters of the network to be fixed, the condition for the second term in the right side of (13) to be positive is:

$$\boldsymbol{s}_w^2 < \frac{2\,th^2\,\boldsymbol{t}^2}{n_m} - w\,th\,\boldsymbol{t} \tag{15}$$

So, increasing the weights standard deviation more than a critical value can increase the speed of pattern propagation as well.

## 4. Simulation results

We evaluated the above discussion through a computer simulation of a two-layered feed-forward network of non-leaky integrate-and-fire neurons. Each layer consists of *100* neurons. As mentioned above, the obtained results for two successive layers can be generalized for a network having an arbitrary number of layers. To be a close approximation to the biological reality, the membrane time constant is set to *20 msec* and the threshold to *20 mV* above the resting potential (McCormik et al 1985). The synaptic delay for each connection is set to be *5 ms*. The membrane potentials of the neurons (eq.(1)) were solved numerically by the forward Euler method. The time step was set to *0.01 ms*.

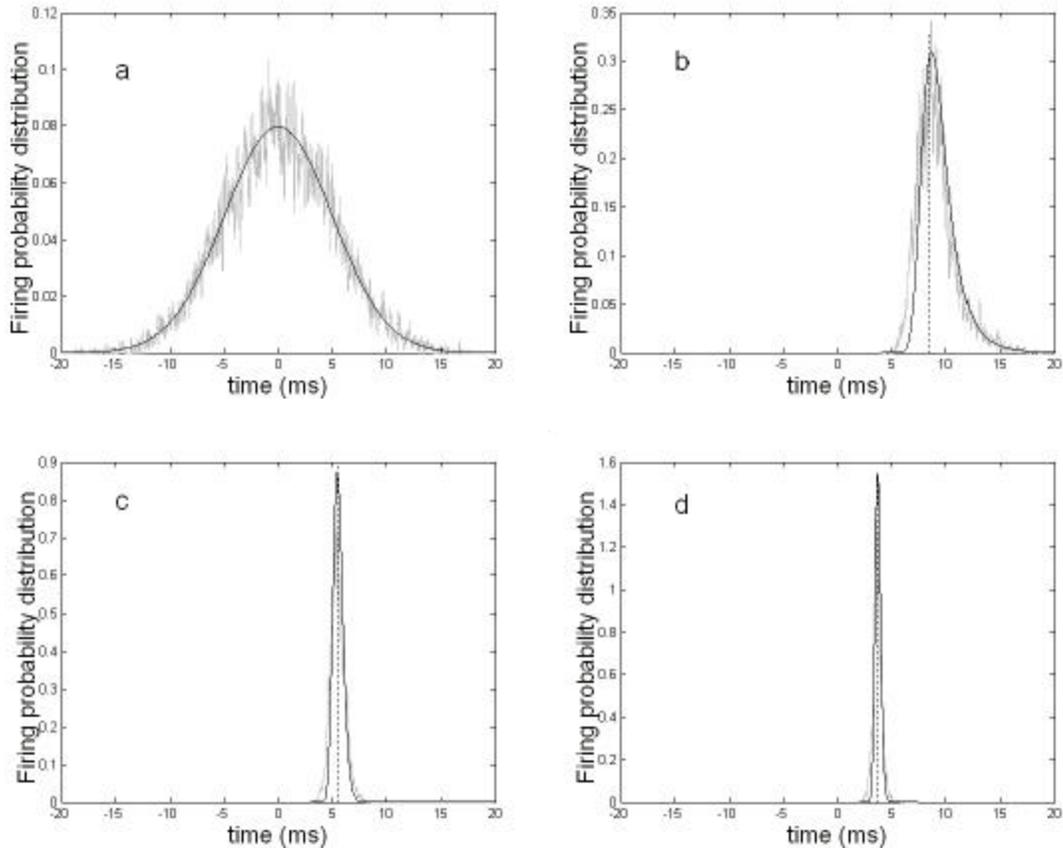

**Figure 2.** The probability density distribution for firing times obtained by simulation (gray curves) and also analytic results (black solid curves) for **a)** the first layer of the network. Note that the neurons in the first layer are set to fire with a mean firing time of *0* and standard deviation of *5 ms*. The solid curve is the result of eq.(2) for these parameters **b)** The probability density distribution for firing times of the neurons in the second layer when $\bar{w}=5$ and $s_w=5$, **c)** when $\bar{w}=7.93$ and $s_w=5$, **d)** when $\bar{w}=10$ and $s_w=5$. The solid curves for the second layer (b-d) are the results of eq. (10) and the dotted lines are the approximated results for the mean firing times from eq(12).

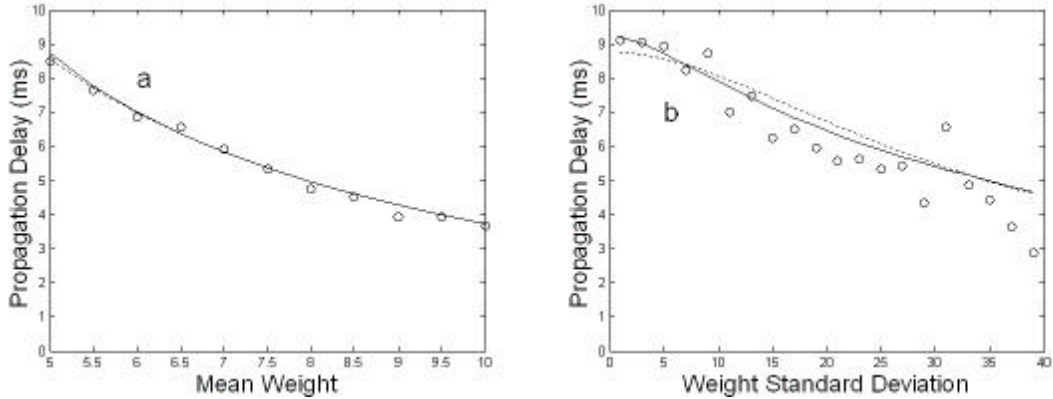

**Figure 3.** The effect of weight parameters on propagation delay. **a)** Propagation delay as a function of mean weight. **b)** Propagation delay as a function of weight standard deviation. The circles show the simulation results, the solid line is the propagation delay numerically calculated by eq.(10) and the dotted line is the result of eq.(13).

During the simulation, we imposed a normal distribution of firing times to the first layer (fig 2.a), and then studied the distribution of firing times in the second layer. The mean-to-mean interval of the two distributions is a measure for pattern propagation delay between the two layers. All the results are the average of *100* realizations of the network with the corresponding parameters

Figure 2 illustrates the density distribution of firing times of the neurons in the first and second layer for a number of weight parameters obtained the simulation (gray curves). The analytic results are also drawn for comparison (solid curves). The firing times of the neurons in first layer were set to obey a normal distribution with mean *0* and standard deviation *5 ms* (figure 2.a). Figures (2.b-2d) show that our analytic formula for the density distribution of firing times in the second layer (eq (10)), which is the cornerstone of our analytic results, satisfactorily feats the simulation results. The approximated mean of the distribution for layer 2 (eq (12)) is also illustrated (dashed lines). Also our approximation for the mean of firing times is acceptable in a wide range of parameters.

As it can be seen, the output spike times in the second layer in turn have a somehow bell-shaped (though not symmetric) density distribution curve and can be approximated by normal distributions, so our analysis can be generalized for more than two layers.

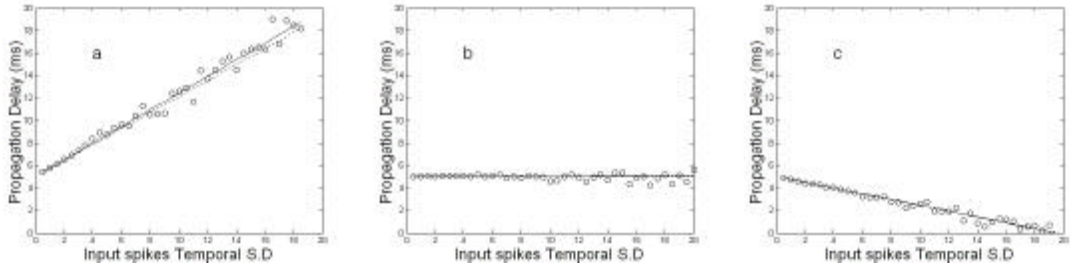

**Figure 4.** Propagation delay as a function of input pattern temporal standard deviation. **a)** When $\bar{w}=5$, **b)** when $\bar{w}=7.93$, **c)** when $\bar{w}=10$. The circles show the simulation results, the solid line is the propagation delay numerically calculated by eq.(10) and the dotted line is the result of eq.(13).

Also note that the distribution of spike times in the second layer is narrower than that of the first layers, i.e. the neurons in the second layers are more synchronized as mentioned by many researchers (Abeles 1991; Herrmann et al 1995; Marsalek 1997; Burkitt & Clrak 1999).

Figure 3.a illustrates the propagation delay between the two layers as a function of mean connection weight, when the other network parameters are fixed and the neurons in the first layer fire with a temporal standard deviation of 5 ms. The simulation results are shown by circles, the exact analytic result derived numerically from eq.(10) is shown by the solid line and the approximated delay from eq.(13) by the dashed line. As expected, by increasing the mean weight, the propagation delay decreases (i.e. the propagation speed increases), so that it can become even shorter than the connection delay. Figure 3.b illustrates the propagation delay as a function connection weights standard deviation, when the other network parameters are fixed. Here also, the propagation delay decrease by increasing the weights standard deviation, as expected.

Figure 4 shows the propagation delay as a function of input spikes temporal standard deviation. In figure 4.a, the mean weight is set to be $\bar{w}=5$. According to eq.(14) we expect the propagation delay to increase by increasing the input spikes temporal standard deviation. As it can be seen, the propagation delay increase almost linearly by increasing the input spikes standard deviation.

In figure 4.b, the mean weight is set to be $\bar{w}=7.93$, which according to eq.(14) must result the propagation delay to be insensitive to the input spikes standard deviation. The simulation results confirm this expectation. Finally, the mean weight is set to be $\overline{w}=10$ in figure 4.c. The simulation results show that, as expected from eq.(14), the propagation delay remains unchanged by increasing the input spikes standard deviation in this case. So increasing the input spikes standard deviation (i.e. the amount of asynchrony in input spikes) can increase, decrease or have no effect on the pattern propagation speed depending on other network parameters.

## 4. Discussion

In summery, we have studied the effect of weight distribution parameters and also the input spikes jitter (temporal standard deviation) on the speed of pattern propagation between successive layers of a synfire chain, by analytical methods and also computer simulation. Increasing the mean or standard deviation of connection weights increases the pattern propagation speed. The input spikes asynchrony can increase or decrease this speed almost linearly, depending on the weight parameters.

When the input spikes are fully synchronous, the propagation speed depends only on the connection delays between the neurons, which are more structural properties of the network compared with connection weights and do not change rapidly during the process of learning and plasticity. This result emphasizes once again the importance of synchronous firing patterns (as proposed in the initial notion of synfire chain) as the mechanism involved in generation of precise firing sequences. When the input spikes to the networks are fully synchronous, the network can generate reproducible firing sequences, which do not change time by time with variations of the synaptic weights. Furthermore, as it can be seen in figure 4, by increasing the temporal standard deviation of input spikes, the random deviation of the propagation delay from its expected value increases. So asynchrony in the input pattern may hinder the generation of reproducible firing sequences, even if the network parameters remain unchanged between two trails.

On the other hand, the linear dependence of the propagation speed on the input spikes temporal standard deviation, shows that a finite amount of asynchrony in the propagating patterns of a feed-forward network, can act as a control parameter for the pattern propagation speed. The biological relevance of this conjecture remains to be studied.